\documentclass[]{aa}
\usepackage{graphicx}
 \usepackage{txfonts}

\begin{document}

   \title{Intra-day variability observations of S5 0716+714 over 4.5 years at 4.8~GHz}

   \author{X. Liu\inst{1,2}
           \and
            H.-G. Song\inst{1,3}
           \and
           N. Marchili\inst{4, 5}
           \and
           B.-R. Liu\inst{1,3,6}
           \and
           J. Liu\inst{1,3}
           \and
         T.P. Krichbaum\inst{4}
         \and
         L. Fuhrmann\inst{4}
         \and
         J.A. Zensus\inst{4}
          }
  \offprints{X. Liu: liux@xao.ac.cn}

\institute{Xinjiang Astronomical Observatory, Chinese Academy of
Sciences, 150 Science 1-Street, Urumqi 830011, PR China \and Key
Laboratory of Radio Astronomy, Chinese Academy of Sciences,
Nanjing 210008, PR China \and Graduate University of the Chinese
Academy of Sciences, Beijing 100049, PR China \and
Max-Plank-Institut f\"ur Radioastronomie, Auf dem H\"ugel 69,
53121 Bonn, Germany \and Dipartimento di Astronomia, Universit\`a
di Padova, Vicolo dellsservatorio 3, 35122, Padova, Italy \and
College of Physical Science and Technology, Guangxi University,
Nanning 530004, Guangxi, PR China
 }

    \date{Received / Accepted }

  \abstract
  {}
  {We aim to search for
evidence of annual modulation in the time scales of the BL Lac
object S5~0716+714.}
   {The intra-day variability (IDV) observations were carried out monthly from 2005 to 2009,
with the Urumqi 25m radio telescope at 4.8~GHz. }
  {The source has shown prominent IDV as well as long-term
flux variations. The IDV time scale does show evidence in favor of
an annual modulation, suggesting that the IDV of 0716+714 is
dominated by interstellar scintillation. The source underwent a
strong outburst phase between mid-2008 and mid-2009; a second
intense flare was observed in late 2009, but no correlation
between the total flux density and the IDV time scale is found,
implying that the flaring state of the source does not have
serious implications for the general characteristics of its
intra-day variability. However, we find that the inner-jet
position angle is changing throughout the years, which could
result in an annual modulation noise in the anisotropic ISS model
fit. There is also an indication that the lowest IDV amplitudes
(rms in flux density) correspond to the slowest time scales of
IDV, which would be consistent with an ISS origin of the IDV of
0716+714. }


   \maketitle

\section{Introduction}

Blazars comprise an extreme subgroup of active galactic nuclei
(AGNs), which consist of flat-spectrum radio quasars and BL Lac
objects. They are variable on time scales ranging from less than a
day to many years. This violent behavior of blazars is generally
explained in terms of a relativistic jet oriented very close to
our line of sight (Urry \& Padovani 1995).

Intra-day variability (IDV) of AGN at centimeter wavelengths was
discovered in the 1980s (Witzel et al. 1986; Heeschen et al.
1987). Today we know that IDV occurs in 25\% -- 50\% of the
flat-spectrum radio sources (Quirrenbach et al. 1992; Lovell et
al. 2008), and 60\% of the bright Fermi blazars (Liu et al. 2012).
Since its discovery, two main explanations for the very short time
scale variability have been proposed, one is that the IDV is
intrinsic to the sources, but this frequently leads to a very high
brightness temperature of the emitting components (Qian et al.
1991) that far exceeds the inverse-Compton limit ($10^{12}$K, see
Kellermann \& Pauliny-Toth 1969). Another explanation is that the
IDV is caused by propagation effects, namely by interstellar
scintillation (ISS) in our galaxy (see Kedziora-Chudczer et al.
1997; Dennett-Thorpe \& de Bruyn 2000; Bignall et al. 2003). For
the ``classical" type-II IDV sources (variability time scales $<$~
0.5--2 days), however, the origin of the variability is not
completely understood (e.g Fuhrmann et al. 2008).

We have carried out a monitoring program for a sample of IDV
sources from August 2005 to January 2010 with the Urumqi 25m radio
telescope at 4.8~GHz. From the analyzed data, at least two IDV
sources in the monitoring program have exhibited systematic
changes of their variability time scales over the year (Gab\'anyi
et al. 2007; Marchili et al. 2012). This effect is known as the
annual modulation of the time scales (e.g., Rickett et al. 2001);
it is explained by assuming that the origin of the variability is
interstellar scintillation. The scattering material is regarded to
be located in a thin plasma screen at a distance on the order of
tens or hundreds of parsecs from the Earth. The orbital motion of
the Earth around the Sun leads to changes in the relative velocity
between the observer and the scattering screen
--- the faster the Earth's movement with respect to the scattering
screen, the shorter the variability time scale. Because the
Earth's velocity follows a one-year periodic cycle, the relative
velocity between the scattering screen and the observer should
change accordingly, resulting in a seasonal cycle of the
variability time scale.

The principal aim of our monitoring program is to search for
evidence of annual modulation in the time scales of type-II IDV
sources. Among the main targets of our monitoring campaign there
is S5~0716+714. It is a BL Lac object; from optical imaging of the
host galaxy, Nilsson et al. (2008) suggested a possible redshift
of 0.31. S5~0716+714 is one of the most variable and compact
blazars, showing multi-wavelength variability from radio to gamma
ray (Raiteri et al. 2003; Abdo et al. 2010). VLA data show a
halo-like jet (Antonucci et al. 1986; Wagner et al. 1996); VLBI
images show a core-dominated jet pointing to the north that is
misaligned with the VLA jets by $\sim90^\circ$ (e.g., Bach et al.
2005). From multi-band long-term monitoring data, radio and
optical light-curve behaviors appear to be quite different, only
minor radio flux enhancements are found simultaneously with the
major optical outbursts (Raiteri et al. 2003). On short time
scales, strong intra-day variability is found in both radio and
optical bands. In 1990, the detection of simultaneous transitions
from fast to slow variability modes among radio and optical
wavelengths during a four-week monitoring campaign suggested a
common, source-intrinsic origin of the variability (see
Quirrenbach et al. 1991). Since then, several multi-frequency
observing campaigns have been carried out for S5~0716+714;
significant evidence in favor of a correlation between optical and
radio variability has not been found anymore, as discussed, e.g.,
in Fuhrmann et al. (2008). These authors hypothesized that both
source-extrinsic and source-intrinsic mechanisms contribute to the
IDV of the source, and that the importance of the two
contributions may depend on the source opacity. In the
source-extrinsic explanation, intra-day variations at frequencies
of 3--8 GHz are attributed to ISS at the border between weak-ISS
and strong-ISS (Rickett 2007). In the regime of weak scattering,
relevant for 0716+714 at $\ga$~5 GHz, the emitting components that
are compact enough to show intrinsic variability on time scales of
a day or less might also show ISS on similar time scales. However,
to separate the source-intrinsic from the source-extrinsic
contribution, it is necessary to carry on IDV monitoring programs
that are sufficiently long to investigate the existence of
possible annual modulation effects in the time scales of the
variability.

\section{Observation and data reduction}

The IDV observations were carried out with the Urumqi 25m radio
telescope, 3-5 days per month, when possible, from Aug. 2005 to
Jan. 2010, with a central frequency of 4800~MHz and a bandwidth of
600~MHz; see Sun et al. (2007) for a description of the observing
system. All observations were performed in `cross-scan' mode, each
scan consists of eight sub-scans in azimuth and elevation over the
source position. This enabled us to check the pointing offsets in
both coordinates. After applying a correction for pointing
offsets, we corrected the measurements for the elevation-dependent
antenna gain and the remaining systematic time-dependent effects
by using several steep spectrum and non-variable secondary
calibrators. Finally, we converted our measurements to absolute
flux density with the frequently observed primary calibrator's
assumed flux densities (3C286, 3C48 and NGC7027). The complete
data calibration procedure guarantees a high level of accuracy, on
the order of 0.5\%, in normal weather conditions.

Following the scheme by Kraus et al. (2003), some quantities were
used to evaluate significance and amplitude of the variability,
namely the reduced chi-square-test, the rms flux density over mean
flux density (the so-called modulation index, $m$), and the
3$\sigma$ relative variability amplitude $Y$, which is corrected
for noise-bias, defined as $Y=3\sqrt{m^{2}-m_{0}^{2}}$, where
$m_{0}$ is the mean modulation index of all calibrators,
describing the statistical measurement accuracy during the
observation. In Table~\ref{tab1}, we list the observational
information and the results of the observations, in which the time
scales are obtained from a structure function analysis (SF) as
introduced in the next section. The columns give; (1) the epoch,
(2) the day of year (DoY); (3) and (4) the duration of observation
and the effective number of data points; (5) and (6) the SF time
scale and relative error; (7), (8) and (9) the modulation index of
calibrators, the modulation index of 0716+714 and the relative
variability amplitude of the source; (10) the source's average
flux density and the rms variation of the flux density; (11) the
reduced $\chi_r^2$.

\begin{table*}
         \caption[]{The observational information and the results derived from the 4.8~GHz observations.}
         $$
         \begin{tabular}{cccccccccccc}


\hline
  \hline
    \noalign{\smallskip}
    1&2 &3 &4 & 5&6 &7 & 8&9 &10 &11  \\
    Start Day & DoY & dur& NP& $t_{SF}$ & error & $m_{0}$ & $m$ & $Y$ & $\overline{S}_{4.8GHz}\pm rms$ & $\chi_r^2$ \\

  & &  (d)& &(d)&  & [\%] & [\%] & [\%] & (Jy) & \\

\hline
  \noalign{\smallskip}

  14.08.2005   & 228    &  2.9    &   24      &  0.7   &  0.3  & 0.8   & 3.3  &  9.6  &  0.880$\pm$0.029  & 7.75 \\

  27.12.2005  & 363     &  3.7    &   55      &  0.6   &  0.2  & 1.2   &  5.3  & 15.6  &  0.823$\pm$0.044  & 23.21 \\

  15.03.2006   & 76     &  3.0    &   50      &  0.8    & 0.2  & 0.5   & 1.9  &  5.4  &  0.638$\pm$0.012  &  2.72 \\

  28.04.2006  & 119    &  3.9    &   70      &  1.1   &  0.2  & 0.5   &  2.0  & 5.9   &  0.644$\pm$0.013  & 4.20 \\
  10.06.2006  & 162     &  3.2    &   89     &  1.3   &  0.3  & 0.5   &  4.9  & 14.6   &  0.735$\pm$0.036  & 24.32 \\

  14.07.2006  & 198     & 4.0     &   87    &  0.5   &  0.2  &  0.6  &  2.7  & 7.8   &  0.748$\pm$0.020    & 6.05 \\

  19.08.2006  & 235     & 2.3      &  67    &  0.9   &  0.3  &  0.5  &  3.0  &  8.8  &  0.845$\pm$0.025  & 13.40 \\

  23.09.2006  & 269     & 5.0      &  141     &  2.1   &  0.3  & 0.5   & 1.8  & 5.3   &  0.814$\pm$0.015   & 4.08 \\

  17.11.2006  & 324      & 4.9     &  133     &  1.1   &  0.3  &  0.5  &  3.2  & 9.6   &  0.745$\pm$0.024 & 9.32 \\

  18.12.2006  & 354    &  2.5      &   77     & 0.6   &  0.2  & 0.5   &   2.0 &  5.8  &  0.702$\pm$0.014  & 3.65 \\

  25.01.2007  & 26     & 2.3      &  66      &  1.0   &  0.2  &  0.4  &  2.2  & 6.4   & 0.786$\pm$0.017  & 6.30 \\

  12.02.2007  & 45     & 4.0     &  109     &  0.6   &  0.3  &  0.4  &  2.3  &  6.6  &  0.755$\pm$0.017  & 5.38 \\

  24.03.2007  & 85     & 2.8     &  72      &  1.3   &  0.3  & 0.5   &  2.2  & 6.4   &  0.735$\pm$0.016  & 5.45 \\

  20.04.2007   & 113   & 3.6     &  78      &  0.8   &  0.3  &  0.6  & 4.3  &  12.8   &  0.743$\pm$0.032 & 15.20 \\
  15.06.2007  & 168    & 2.4     &  58      &  0.6   &  0.3  &  0.6  & 2.3  &  6.6  &  0.834$\pm$0.019   & 4.85 \\

  19.07.2007  & 202    & 2.9     &  69     &  1.1   &  0.2  & 0.6   &  4.7  & 14.0   & 0.772$\pm$0.036  & 19.76 \\

  18.08.2007  & 232    & 3.1      &  72     &  1.2   &  0.3  &  0.6  &  4.1  & 12.2   & 0.779$\pm$0.032  & 14.30 \\
  13.10.2007  & 288   & 3.0      &  65     &  1.3   &  0.3  &  0.4  &  2.6  & 7.7   & 0.806$\pm$0.021   & 8.32 \\

  21.12.2007  & 357   & 3.2     &  80    &   1.2   & 0.3  & 0.4   &  2.9  & 8.6   & 0.690$\pm$0.020    & 10.24 \\
  25.02.2008  & 57    & 2.9     &  59    &  0.5   &  0.2  &  0.6  &  2.1  &  6.0  & 0.818$\pm$0.017  & 4.68 \\

  21.03.2008  & 82    & 3.0     &  76     &  0.8   &  0.3  &  0.4  &  2.9  &  8.7  & 0.790$\pm$0.023  & 10.52 \\

  21.04.2008  & 113    &3.1     &  70    &  0.8   &  0.3  &  0.5  &  3.5  &  10.4  & 0.858$\pm$0.030  & 13.06 \\
  21.06.2008  & 174    & 3.5      &  55    &   0.9   &  0.3  &   0.5 &  2.9  & 8.7   & 0.985$\pm$0.029  & 12.57 \\

  18.07.2008  & 202    & 4.8       & 55    &  0.8   &  0.2  &   0.6 &  2.5 & 7.3   &  1.272$\pm$0.032  & 4.75 \\

  20.08.2008  & 235    & 5.0      & 72     &   0.4  &  0.2  &   0.6 &  2.0  & 5.7   & 1.308$\pm$0.026  & 4.56 \\

  12.09.2008  & 258    & 3.6      & 85     &   1.3  &  0.3  &   0.4 &  2.4  & 7.2  & 1.148$\pm$0.028   & 7.04 \\

  06.11.2008  & 313   & 3.6       & 55    &   0.6   &  0.3  &   0.6 &  3.0  & 8.8   &  1.368$\pm$0.041 & 6.10 \\

  22.12.2008  & 358   & 2.3       & 57    &  1.1  &  0.3  &   0.4 &  1.8  &  5.1  &  1.369$\pm$0.024   & 3.04 \\
  11.01.2009  & 12    & 2.6      &  69    &  0.8   & 0.3  &   0.4 &  1.6  &  4.7  &  1.607$\pm$0.026   & 2.51 \\

  23.02.2009  & 55    & 3.0       & 153    &   0.5  &  0.2  &   0.4 &  1.2  &  3.2  &  1.473$\pm$0.017   & 2.23 \\

  21.03.2009  & 82    & 4.9     & 124      &  0.9  &  0.3  &   0.5 &  1.8  & 5.1   &  1.243$\pm$0.022  & 3.38 \\

  19.04.2009  & 112   & 5.4    & 94       & 2.0   &  0.4  &  0.6 &  3.7  &  10.9  &  1.362$\pm$0.050  & 14.02 \\

  06.05.2009  & 128   & 3.9    & 90      & 1.4    & 0.3  &   0.7 &  3.4  & 10.1  &   1.137$\pm$0.039  & 8.37 \\

  25.06.2009  & 177   & 2.6   & 52        & 1.0   &  0.3  &  0.6 &  3.5  & 10.4   &  0.938$\pm$0.033  & 12.27 \\

  21.08.2009  & 235   & 4.1   & 94         & 1.7    & 0.4  &  0.5 &  2.3  &  6.6  &  0.932$\pm$0.021  & 4.96  \\

  22.09.2009  & 268   & 5.5  & 131        & 3.5    &  0.8  &  0.6 &  2.0  & 5.6  &  0.816$\pm$0.016  & 2.86  \\

  09.10.2009  & 283   & 2.3   & 58        & 0.6    &  0.4  &   0.4 &  1.1 &  3.0  & 1.009$\pm$0.011  & 2.02  \\
  22.11.2009  & 328   & 3.8   & 76        & 1.3    & 0.3  &   0.7 &  3.0  & 8.6   & 1.288$\pm$0.038  & 5.30  \\

  11.12.2009  & 347   & 4.4   & 118        & 0.7    & 0.3  &   0.5 &  2.0  & 5.9   &  1.471$\pm$0.030  & 3.57 \\

  19.01.2010  & 21    & 3.5   & 67      &  1.4    &  0.3  &   0.6 &  1.8  &  5.1  &   1.328$\pm$0.024    & 2.60  \\

           \noalign{\smallskip}
            \hline
           \end{tabular}{}
         $$
         \label{tab1}
   \end{table*}

\begin{figure}
     \includegraphics[width=8.5cm]{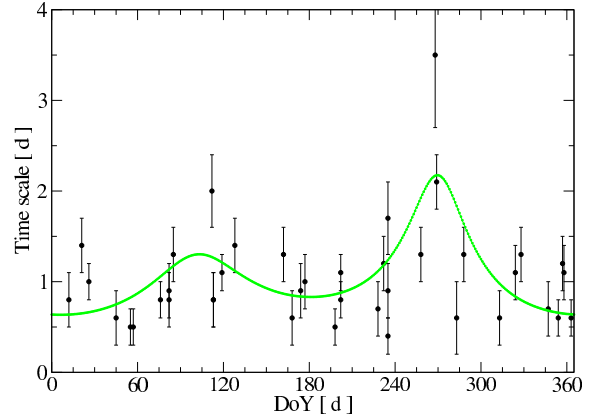}
     \caption{Annual modulation plot of 0716+714. The green line shows the annual modulation pattern that best fits the time scales.}
      \label{fig1}
   \end{figure}

   \begin{figure}
     \includegraphics[width=8.5cm]{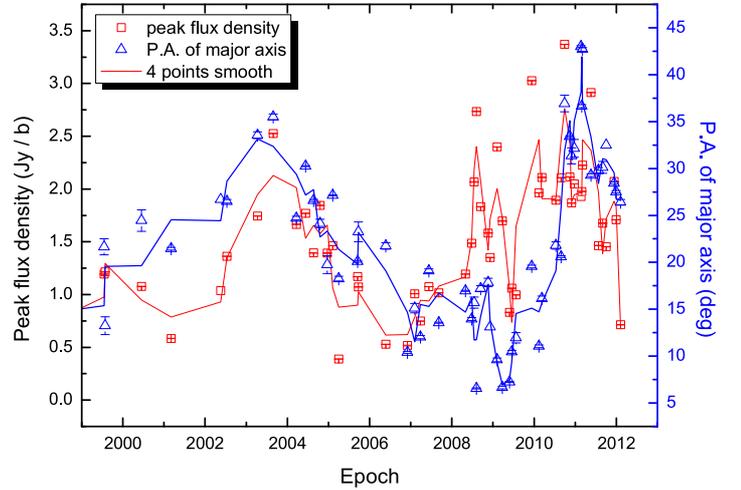}
     \caption{Peak flux density (per beam) and the position angle of the VLBA `core' at 15~GHz versus epoch of
     observation.}
      \label{fig2}
\end{figure}

    \begin{figure}
     \includegraphics[width=7.5cm]{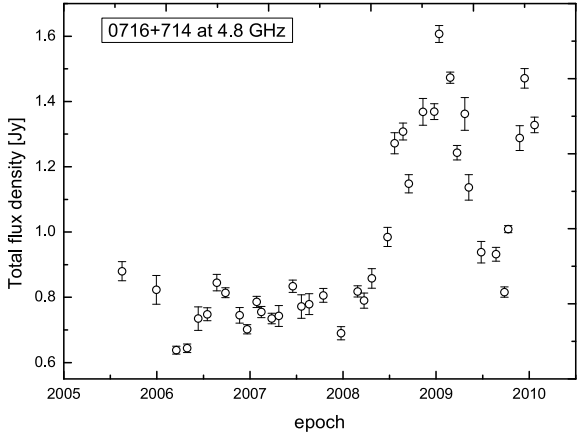}
     \caption{Total flux density versus epoch of observation at 4.8~GHz .}
      \label{fig3}
   \end{figure}

   \begin{figure}
     \includegraphics[width=7.5cm]{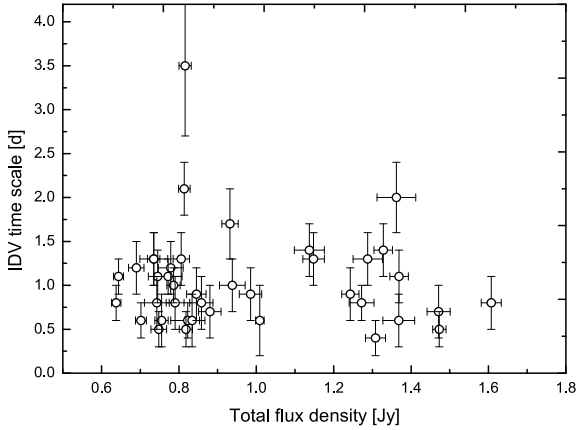}
     \caption{IDV time scale versus total flux density S.}
      \label{fig4}
   \end{figure}

   \begin{figure}
     \includegraphics[width=7.5cm]{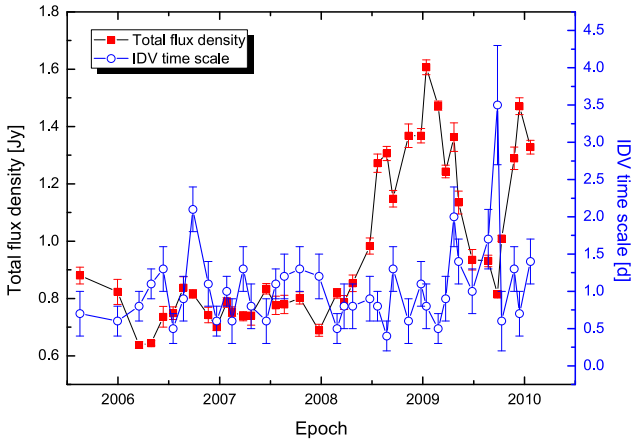}
     \caption{Total flux density and the IDV time scale versus epoch of
     observation.}
      \label{fig5}
   \end{figure}

   \begin{figure}
     \includegraphics[width=7.5cm]{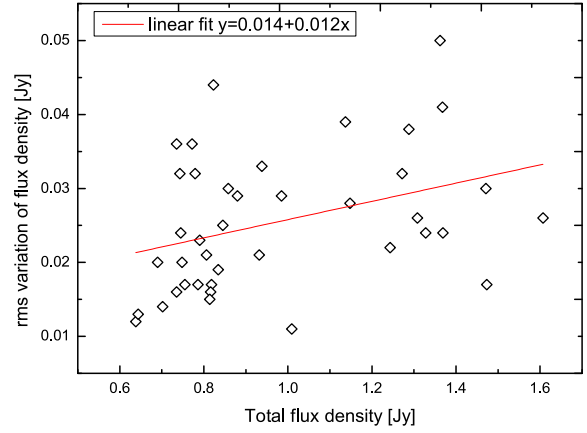}
     \caption{rms flux density variation of S versus average S.}
      \label{fig6}
   \end{figure}

   \begin{figure}
     \includegraphics[width=7.5cm]{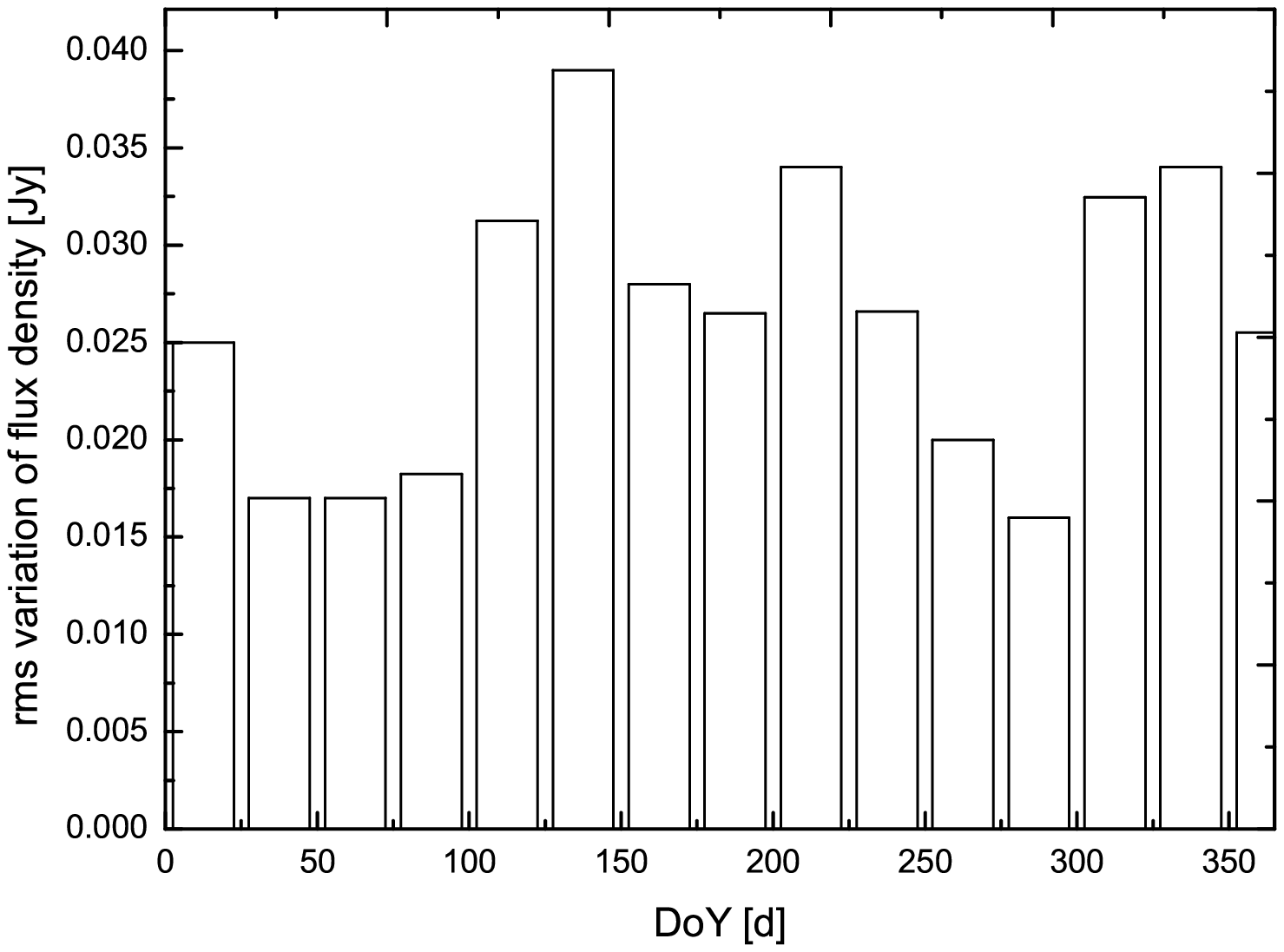}
     \caption{25-day-bin-averaged rms flux density variation of S versus day of
     year.}
      \label{fig7}
   \end{figure}

\section{Variability analysis and discussion}

From the results of the IDV observations in Table~\ref{tab1}, and
according to a $\chi^{2}$ test, 0716+714 exhibits prominent IDV in
all observing sessions at a confidence level of $ \geq 99.9$\,\%.
Here, as a criterion for the source variability, the hypothesis of
a constant function is examined; the datasets with a probability
to be constant $\leq$\,0.1\% are considered to be variable.

To analyze the variability time scales, we used the standard
structure function method (SF), i.e. a first-order structure
function analysis (Simonetti et al. 1985). Above the noise level,
ideally, the SF rises monotonically with a power law shape and
reaches its maximum at a `saturation' level. The intersection of
the power law fit with the plateau corresponding to this
saturation level defines the characteristic variability time
scale. In fact, the plateau is often not well pronounced, but it
can be estimated by the mean of the SF around the first maximum.
The errors of the power law fit to the SF have also to be taken
into account. Depending on the uncertainties in the evaluation of
both the SF saturation level and the power law fit, the estimated
characteristic time scale changes. The error on the estimation of
the time scale can therefore be obtained by taking into account
the formal errors of the power-law fit and the fit to the SF
plateau. Sometimes, the structure function may show more than one
plateau, indicating the existence of multiple variability time
scales. In those cases, we identified the characteristic time
scale with the shortest one (Marchili et al. 2012).

In Fig.~\ref{fig1} we plot the variability time scales versus the
day of year (the so-called annual modulation plot) for all
observing sessions. To investigate the possible existence of an
annual modulation in the time scales of 0716+714, we fitted the
time scales according to the model described in Qian \& Zhang
(2001), updated to take into account the case of anisotropic
scattering (see, e.g., Bignall et al. 2006; Gab\'anyi et al.
2007). An anisotropic scattering can be caused by either an
elongation of the scintles (i.e., patches of focused or defocused
light across which the Earth is moving) in a given direction, or
an anisotropy of the emitting component (e.g., Gab\'anyi et al.
2009).

According to our model (for more details of the ISS model-fit
code, see Marchili et al. 2012) the variability time scale is
expressed as a function of the day of year, and depends on the
orientation of the elliptical scintillation pattern (a unit vector
${\bf s}=(\mathrm{cos\theta, sin\theta})$), the relative velocity
between the scattering screen and the Earth, $\bf
v\mathrm{(DoY)}=\bf v_{\mathrm{ISS}}-\bf
v_{\oplus}\mathrm{(DoY)}$, the distance to the screen, D, and the
anisotropy factor, r:

\begin{equation}
t(\mathrm{DoY}) \propto \frac{\mathrm{D}\cdot
\sqrt{\mathrm{r}}}{\sqrt{{\mathrm{v}^2(\mathrm{DoY})+(\mathrm{r}^2-1)\,({\bf
v(\mathrm{DoY}) \times s})^2}}}.
\end{equation}

The algorithm for the least-squares fitting of the time scales
uses five free parameters: the relative velocity ${\bf v}$,
projected onto the right ascension and the declination coordinates
(which allows one to fit the screen velocity ($\bf
v_{\mathrm{ISS},\alpha}$ and $\bf v_{\mathrm{ISS},\delta}$)
relative to the Local Standard of Rest, since the Earth velocity
is known with respect to the LSR), the distance to the screen, the
anisotropy degree and the anisotropy angle $\theta$ (measured from
east through north), which is derived from the vectorial product
${\bf v(\mathrm{DoY}) \times s}$. The parameters that best fit the
time scales of 0716+714 are reported in Table~\ref{tab2}.

The result with the best fit anisotropic screen model is shown in
Fig.~\ref{fig1}; there is significant evidence in favor of an
annual modulation of the time scales, which exhibit a remarkable
slow-down peaking around DoY 270 and a secondary peak occurring
around DoY 100. The screen appears to be slightly anisotropic,
with an anisotropy ratio of about 1.7 and an anisotropy angle
about 80 degrees. To follow the usual convention of radio image
analysis, the position angle of the anisotropy is
$90\degr-\theta=10\degr$ from north to east, this is roughly
consistent with the inner-jet position angle ranging from a few to
about 35 degrees in the VLBI images of 0716+714 (see, Britzen et
al. 2009). Therefore the anisotropy might be also caused by an
anisotropy of the emitting component in 0716+714. However, the
data do not allow us to ambiguously determine whether the
anisotropy originates from the intrinsic source structure or from
the scattering screen. According to our model, the variability is
associated to an interstellar cloud between the Earth and 0716+714
at a distance of 230 pc, whose characteristics could be
investigated in the future, as was done for some other ISS-induced
objects (Linsky, Rickett \& Redfield 2008).

Anisotropic ISS models have been applied to several other IDV
objects showing an annual modulation pattern, e.g., J1819+3845
(Dennett-Thorpe \& de Bruyn 2003), PKS 1257$-$326 (Bignall et al.
2006), J1128+592 (Gab\'anyi et al. 2009), PKS 1519$-$273 and PKS
1622$-$253 (Carter et al. 2009), and S4 0954+65 (Marchili et al.
2012). It is challenging to detect an annual modulation of the
variability time scales in IDV sources (in particular in the
slower type-II sources) -- a large amount of long (several days)
IDV observations over years have to be performed. For our project,
the observations were often not evenly and densely allocated over
time; one has to overlap the data of years into the day of year
(DoY); only after several years of observations, we were able to
detect and fit an anisotropic seasonal cycle as shown in
Fig.~\ref{fig1}. The modeling assumes, however, that the ISS
scattering screen is stable over several years. This assumption,
however, is not necessarily true, as shown, e.g., in the case of
J1819+3845, where the scattering medium that is responsible for
the strong and rapid IDV of the source has moved away, leading to
a significant decrease of the variability (Macquart \& de Bruyn
2007; Koay et al. 2011). Changes in the scattering screen
throughout the years, for instance, changes of the turbulent
patches in the ISM (e.g. changes in the scattering measure,
distance and/or anisotropy) will mostly result in changes of the
scintillation strength, but may also affect and reduce the
significance of the time scale fitting with an ISS model. This
would explain the relatively high $\chi_{r}^{2}$ values frequently
found for the best model fits of several IDV sources, such as
J1819+3845 ($\chi_{r}^{2}$=1.5; Dennett-Thorpe \& de Bruyn 2003),
PKS 1257$-$326 ($\chi_{r}^{2}$=1.97; Bignall et al. 2006), PKS
1519$-$273, and PKS 1622$-$253 ($\chi_{r}^{2}$=0.8 and 2.1,
respectively; Carter et al. 2009), and J1128+592
($\chi_{r}^{2}$=3.0; Gab\'anyi et al. 2007). For 0716+714, we
found a $\chi_{r}^{2}$ of 2.5, which is comparable with the
results reported for the sources mentioned above. On the other
hand, considering that the position angle of the 0716+714 jet is
close to the anisotropy angle derived from our ISS model fit, it
is plausible that the anisotropic scattering is caused by a
source-intrinsic anisotropy. Because the inner-jet position angle
is oscillating from $\sim10{\degr}$ to $\sim35{\degr}$ following a
5.7$\pm$0.5-year long-term variability cycle in the total flux
density of 0716+714 (Raiteri et al. 2003; Fan et al. 2007; Britzen
et al. 2009), the anisotropy angle may vary accordingly. This
would also contribute to an increase of the $\chi_{r}^{2}$ of the
annual modulation fit.

We have model-fitted the core (inner-jet) of the 15 GHz MOJAVE
images (Lister et al. 2009) of 0716+714 and obtained the peak flux
density (per beam) and the position angle evolution over 12 years;
as shown in Fig.~\ref{fig2}, the position angle (PA) positively
correlates with the peak flux density, resulting a linear Pearson
correlation coefficient of 0.44 (significance 4.6E-4). During our
IDV observations in 2006-2009, the PA first decreased, then
increased following the same trend of the flux density, and then
decreased, with a PA variation of about 10$\degr$. If the
scintillating component in the source is anisotropic and
contributes to the anisotropy of the ISS scattering, the ISS model
fit to the 4.5 years of collected IDV time scales should be
considerably affected, with a significant increase of the
$\chi_{r}^{2}$. The inner-jet PA evolution would influence the IDV
time scales year by year; to investigate this effect, more densely
sampled IDV observations and careful year-by-year anisotropic
modeling would be necessary. In our case, the IDV time scale data
are still too sparse in every single year to model a change of
anisotropic scattering pattern induced by the source's PA
evolution.

In Table~\ref{tab2}, the screen velocity from our model is much
lower than the Earth orbiting velocity w.r.t. the LSR, indicating
that the Sun's motion plays the main role in the variation of the
time scales. For 0716+714's position on the sky, the time scale
peak falls around DoY 270 (Fig.~\ref{fig1}) as expected,
supporting the hypothesis that ISS is the dominating contribution
to the IDV of 0716+714.

During the 4.5 years of monitoring, the flux density of 0716+714
showed strong variations also on time scales of months
(Fig.~\ref{fig3}); a flare appears from mid-2008 to mid-2009 with
peak-to-through variations on the order of 100\%, and a second
flare occurs late in 2009. This long-term flux density variation
should have a source-intrinsic origin. The IDV time scales during
the flaring state could be prolonged due to, e.g., an enlargement
of the scintillating component. However, in Table~\ref{tab1} and
Fig.~\ref{fig4} we see that the IDV time scales in the flaring
state (e.g. $>$ 1 Jy) are not very different compared to those
estimated during the relatively `quiescent' state (e.g. $<$ 1 Jy)
in general, and no correlation is found between the IDV time scale
and the total flux density, implying that the source flares do not
seriously affect the variability time scales of 0716+714. The
total flux density and the IDV time scale are plotted versus the
observing epoch in Fig.~\ref{fig5}. There is no correlation
between the two, however, we can not completely rule out that the
slower time scales observed in 2009 are somehow related to the
2008 flare taking into account a time delay of about one year. For
the inner-jet kinematics of 0716+714, Britzen et al. (2009)
proposed a model in which the VLBI components of 0716+714 are
stationary with respect to the core, while the inner components
are oscillating with regard to their PA. In this model, the flare
in flux density is just caused by a geometric beaming effect
--- the inner components of 0716+714 do not change physically from
a quiescent state (less beaming) to a flaring state (strong
beaming), unlike the PA, which instead changes considerably; as a
result, the projected size and/or anisotropy of the scintillating
component onto our line of sight must change over the years. It is
hypothesized that the inner-jet PA evolution of 0716+714 follows a
$\sim$5.7-year modulation, which would affect the annual
modulation of IDV time scales from year to year; future densely
sampled IDV observations, e.g. every week, might be able to detect
such an effect.

In the ISS-induced variability, the root mean square of the flux
density is expected to increase linearly with the average flux
density (Narayan 1992). Plotting the two parameters against each
other (Fig.~\ref{fig6}) indicates that there seems to be a weak
positive correlation between them, with a linear Pearson
correlation coefficient of 0.36 (significance of 0.02). This weak
correlation may be explained either by an increase of the
contribution of instrumental noise to the overall variability at
the lowest flux densities, or by an rms dependence on the
variability time scale
---  when this becomes longer than the observation duration, part
of the variability may fall outside the observing window, leading
to a decrease of the rms. If this is the case, we should see a
decrease of the rms around the time of the year when the
variability is slower. Plotting the rms flux density (with bins of
25 days) versus the day of year (Fig.~\ref{fig7}), we find a
trough around DoY 275 and a secondary trough around DoY 70. The
result suggests that the IDV amplitude is low at the slowest time
scales observed in the annual modulation plot (Fig.~\ref{fig1});
the trough of the rms flux density around DoY 70 only poorly
coincides with the secondary peak of time scale around DoY 100,
but the difference is moderate.

 \begin{table}
         \caption[]{The best fit screen parameters from the IDV time scales of 0716+714 at 4.8~GHz.
}
         $$
         \begin{tabular}{ccccc}
            \hline
            \noalign{\smallskip}

$v_{ISS,\alpha}$ & $v_{ISS,\delta}$ & Screen & Anisotropy & Anisotropy  \\
 to LSR & to LSR &   distance  & degree & angle \\
   (km/s)& (km/s) & D(kpc) & r(ratio) & $\theta(E \rightarrow N)$  \\

            \noalign{\smallskip}
            \hline
            \noalign{\smallskip}

 $1\pm4$  &$10\pm5$ & $0.23\pm0.05$ & $1.7\pm0.4$ & $80\degr\pm15\degr$     \\

           \noalign{\smallskip}
            \hline
           \end{tabular}{}
         $$
         \label{tab2}
   \end{table}

\section{Summary}

We have carried out monthly IDV observations of the blazar
0716+714 over 4.5 years with the Urumqi 25m radio telescope at
4.8~GHz; the source has shown prominent IDV as well as long-term
flux variations. With the structure function analysis we found
that the IDV time scale does show evidence in favor of a seasonal
cycle, a result which suggests that the IDV of 0716+714 is caused
by interstellar scintillation. The source underwent a strong
outburst phase between mid-2008 and mid-2009; a second intense
flare was observed in late 2009, but no correlation between the
total flux density and the IDV time scale is found, implying that
the flaring state of the source does not have serious implications
for the general characteristics of its intra-day variability.
However, we know that the inner-jet position angle is changing
during the years, which could result in a significant variation of
the annual modulation pattern with time, and therefore a decrease
in the significance of the anisotropic ISS model fit to the IDV
time scales. We also found indications that the lowest IDV
amplitudes (rms flux density) correspond to the slowest time
scales of the variability, which we were able to explain
reasonably well within the ISS model.

\begin{acknowledgements}
We thank the anonymous referee for valuable comments, which have
improved the paper. This research has made use of data from the
MOJAVE database that is maintained by the MOJAVE team (Lister et
al., 2009, AJ, 137, 3718). This work is supported by the National
Natural Science Foundation of China (Grant No.11073036) and the
973 Program of China (2009CB824800). N.M. is funded by an ASI
fellowship under contract number I/005/11/0.
\end{acknowledgements}

\object{S5 0716+71}
\object{[HB89] 0716+714}

\end{document}